\documentclass{article}

\def\abstract#1{\vskip 7mm 
        \begin{center}{\large Abstract}\par \smallskip
                \begin{minipage}[c]{12cm}
                        \small #1
                \end{minipage}
        \end{center}
}
\def\title#1{\begin{center}{\Large\bf #1}\end{center}}
\def\author#1{\vskip 5mm \begin{center}{#1}\end{center}}
\def\address#1{\begin{center}{\it #1}\end{center}}
\makeatletter
\@ifundefined{lesssim}{}{}
\@ifundefined{gtrsim}{}{}
\def\vereq#1#2{\lower3pt\vbox{\baselineskip1.5pt \lineskip1.5pt
\ialign{$\m@th#1\hfill##\hfil$\crcr#2\crcr\sim\crcr}}}
\makeatother

\begin{document}

\title{%
  Consistency relations between the source terms in the
  second-order Einstein equations for cosmological perturbations
}
\author{%
  Kouji NAKAMURA\footnote{E-mail:kouchan@th.nao.ac.jp},
}
\address{%
  Department of Astronomical Science,
  the Graduate University for Advanced Studies,\\
  Mitaka, Tokyo 181-8588, Japan.
}

\abstract{
  In addition to the second-order Einstein equations on
  four-dimensional homogeneous isotropic background universe 
  filled with the single perfect fluid, we also derived the
  second-order perturbations of the continuity equation and the
  Euler equation for a perfect fluid in gauge-invariant manner
  without ignoring any mode of perturbations. 
  The consistency of all equations of the second-order Einstein
  equation and the equations of motion for matter fields is
  confirmed.
  Due to this consistency check, we may say that the set of all
  equations of the second-order are self-consistent and they are
  correct in this sense.
}


\section{Introduction}
\label{sec:intro}


The general relativistic second-order cosmological perturbation
theory is one of topical subjects in the recent cosmology. 
By the recent observation\cite{WMAP}, the first order
approximation of the fluctuations of our universe from a
homogeneous isotropic one was revealed.
The observational results also suggest that the fluctuations of
our universe are adiabatic and Gaussian at least in the first
order approximation.
We are now on the stage to discuss the deviation from this first
order approximation from the
observational\cite{Non-Gaussianity-observation-WMAP} and the 
theoretical
side\cite{Non-Gaussianity-inflation-Non-Gaussianity-in-CMB}
through the non-Gaussianity, the non-adiabaticity, and so on.
To carry out this, some analyses beyond linear order are
required.
The second-order cosmological perturbation theory is one of
such perturbation theories beyond linear order.


In this article, we confirm the consistency of all equations of
the second-order Einstein equation and the equations of motion
for matter fields, which are derived in
Refs.~\cite{kouchan-JGRG17-proceedigns,kouchan-second-cosmo-matter}.
Since the Einstein equations include the equation of motion for
matter fields, the second-order perturbations of the equations
of motion for matter fields are not independent equations of the
second-order perturbation of the Einstein equations.
Through this fact, we can check whether the derived equations
of the second order are self-consistent or not.
This confirmation implies that the all derived equations of the
second order are self-consistent and these equations are
correct in this sense.


\section{Metric perturbations}
\label{sec:Second-order-cosmological-perturbatios}


The background spacetime for the cosmological perturbations is a
homogeneous isotropic background spacetime.
The background metric is given by
\begin{eqnarray}
  \label{eq:background-metric}
  g_{ab} = a^{2}\left\{
    - (d\eta)_{a}(d\eta)_{b}
    + \gamma_{ij} (dx^{i})_{a} (dx^{j})_{b}
  \right\},
\end{eqnarray}
where $\gamma_{ab} := \gamma_{ij} (dx^{i})_{a} (dx^{j})_{b}$ is
the metric on the maximally symmetric three-space and the
indices $i,j,k,...$ for the spatial components run from 1 to 3.
On this background spacetime, we consider the perturbative
expansion of the metric as
$\bar{g}_{ab} = g_{ab} + \lambda {}_{{\cal X}}\!h_{ab}
+ \frac{\lambda^{2}}{2} {}_{{\cal X}}\!l_{ab}
+ O(\lambda^{3})$,
where $\lambda$ is the infinitesimal parameter for perturbation
and $h_{ab}$ and $l_{ab}$ are the first- and the second-order
metric perturbations, respectively. 
As shown in Ref.~\cite{kouchan-second-cosmo}, the metric
perturbations $h_{ab}$ and $l_{ab}$ are decomposed as
\begin{eqnarray}
  h_{ab} =: {\cal H}_{ab} + {\pounds}_{X}g_{ab},
  \label{eq:linear-metric-decomp}
  \quad
  \label{eq:second-metric-decomp}
  l_{ab}
  =:
  {\cal L}_{ab} + 2 {\pounds}_{X} h_{ab}
  + \left(
      {\pounds}_{Y}
    - {\pounds}_{X}^{2}
  \right)
  g_{ab},
\end{eqnarray}
where ${\cal H}_{ab}$ and ${\cal L}_{ab}$ are the
gauge-invariant parts of $h_{ab}$ and $l_{ab}$, respectively.
The components of ${\cal H}_{ab}$ and ${\cal L}_{ab}$ can be
chosen so that 
\begin{eqnarray}
  \label{eq:components-calHab}
  {\cal H}_{ab}
  &=& 
  a^{2} \left\{
    - 2 \stackrel{(1)}{\Phi} (d\eta)_{a}(d\eta)_{b}
    + 2 \stackrel{(1)}{\nu}_{i} (d\eta)_{(a}(dx^{i})_{b)}
    + \left( - 2 \stackrel{(1)}{\Psi} \gamma_{ij} 
      + \stackrel{(1)}{\chi}_{ij} \right)
    (dx^{i})_{a}(dx^{j})_{b}
  \right\},
  \\
  \label{eq:components-calLab}
  {\cal L}_{ab}
  &=& 
  a^{2} \left\{
    - 2 \stackrel{(2)}{\Phi} (d\eta)_{a}(d\eta)_{b}
    + 2 \stackrel{(2)}{\nu}_{i} (d\eta)_{(a}(dx^{i})_{b)}
    + \left( - 2 \stackrel{(2)}{\Psi} \gamma_{ij} 
      + \stackrel{(2)}{\chi}_{ij} \right)
    (dx^{i})_{a}(dx^{j})_{b}
  \right\}.
\end{eqnarray}
In Eqs.~(\ref{eq:components-calHab}) and
(\ref{eq:components-calLab}), the vector-mode
$\stackrel{(p)}{\nu}_{i}$ and the tensor-mode
$\stackrel{(p)}{\chi_{ij}}$ ($p=1,2$) satisfy the properties 
\begin{eqnarray}
  D^{i}\stackrel{(p)}{\nu}_{i} =
  \gamma^{ij}D_{p}\stackrel{(p)}{\nu}_{j} = 0, \quad
  \stackrel{(p)}{\chi^{i}_{\;\;i}} = 0, \quad
   D^{i}\stackrel{(p)}{\chi}_{ij} = 0,
\end{eqnarray}
where $\gamma^{kj}$ is the inverse of the metric $\gamma_{ij}$.


\section{Background, First-, and Second-order Einstein equations}
\label{sec:Einstein-equations}


The Einstein equations of the background, the first order, and
the second order on the above four-dimensional homogeneous
isotropic universe are summarized as follows.


The Einstein equations for this background spacetime filled with
a perfect fluid are given by 
\begin{eqnarray}
  \label{eq:background-Einstein-equation-1}
  \stackrel{(0)}{{}^{(p)}E_{(1)}}
  := {\cal H}^{2} + K - \frac{8 \pi G}{3} a^{2} \epsilon = 0
  , \quad 
  \stackrel{(0)}{{}^{(p)}E_{(2)}}
  := 2 \partial_{\eta}{\cal H} + {\cal H}^{2} + K + 8 \pi G a^{2}p
  = 0
  ,
\end{eqnarray}
where ${\cal H} = \partial_{\eta}a/a$, $K$ is the curvature
constant of the maximally symmetric three-space, $\epsilon$ and
$p$ are energy density and pressure, respectively.


On the other hand, the second-order perturbations of the
Einstein equation are summarized as
\begin{eqnarray}
  \stackrel{(2)}{{}^{(p)}E_{(1)}}
  &:=&
  \left(
    - 3 {\cal H} \partial_{\eta}
    +   \Delta
    + 3 K
  \right) \stackrel{(2)}{\Psi}
  - 3 {\cal H}^{2} \stackrel{(2)}{\Phi}
  - 4\pi G a^{2} \stackrel{(2)}{{\cal E}}
  - \Gamma_{0}
  = 0
  ,
  \label{eq:kouchan-19.120}
  \\
  \stackrel{(2)}{{}^{(p)}E_{(2)}}
  &:=&
  \left(
                  \partial_{\eta}^{2} 
    +          2  {\cal H} \partial_{\eta}
    -             K
    - \frac{1}{3} \Delta
  \right)
  \stackrel{(2)}{\Psi}
  + \left(
                  {\cal H} \partial_{\eta}
    +          2  \partial_{\eta}{\cal H}
    +             {\cal H}^{2}
    + \frac{1}{3} \Delta
  \right)
  \stackrel{(2)}{\Phi}
  \nonumber\\
  && \quad\quad
  - 4 \pi G a^{2} \stackrel{(2)}{{\cal P}}
  - \frac{1}{6} \Gamma_{k}^{\;\;k}
  = 0
  ,
  \label{eq:kouchan-19.121}
  \\
  \stackrel{(2)}{{}^{(p)}E_{(3)}}
  &:=&
  \stackrel{(2)}{\Psi} - \stackrel{(2)}{\Phi}
  -
  \frac{3}{2}
  \left( \Delta + 3 K \right)^{-1}
  \left(
    \Delta^{-1} D^{i}D^{j}\Gamma_{ij}
    - \frac{1}{3} \Gamma_{k}^{\;\;k}
  \right)
  = 0
  ,
  \label{eq:kouchan-19.122}
  \\
  \stackrel{(2)}{{}^{(p)}E_{(4)i}}
  &:=&
                \partial_{\eta}D_{i}\stackrel{(2)}{\Psi}
  +             {\cal H} D_{i}\stackrel{(2)}{\Phi}
  - \frac{1}{2} D_{i}\Delta^{-1}D^{k}\Gamma_{k}
  +          4  \pi G a^{2} (\epsilon + p) D_{i}\stackrel{(2)}{v} 
  = 0
  ,
  \label{eq:kouchan-19.125}
  \\
  \stackrel{(2)}{{}^{(p)}E_{(5)i}}
  &:=&
  \left( \Delta + 2 K \right) \stackrel{(2)}{\nu_{i}}
  + 2 \left(\Gamma_{i} - D_{i}\Delta^{-1}D^{k}\Gamma_{k}\right)
  - 16 \pi G a^{2} (\epsilon + p) \stackrel{(2)}{{\cal V}_{i}}
  = 0
  ,
  \label{eq:kouchan-19.126}
  \\
  \stackrel{(2)}{{}^{(p)}E_{(6)i}}
  &:=&
  \partial_{\eta}\left(
    a^{2}  \stackrel{(2)}{\nu_{i}}
  \right)
  - 2 a^{2} \left(\Delta+2K\right)^{-1} \left\{
    D_{i}\Delta^{-1}D^{k}D^{l}\Gamma_{kl} - D^{k}\Gamma_{ik}
  \right\}
  = 0
  ,
  \label{eq:kouchan-19.123}
  \\
  \stackrel{(2)}{{}^{(p)}E_{(7)ij}}
  &:=&
  \left(
    \partial_{\eta}^{2}
    + 2 {\cal H} \partial_{\eta}
    + 2 K
    - \Delta
  \right) \stackrel{(2)}{\chi_{ij}}
  - 2 \Gamma_{ij}
  + \frac{2}{3} \gamma_{ij} \Gamma_{k}^{\;\;k}
  \nonumber\\
  &&
  + 3
  \left(
    D_{i}D_{j} - \frac{1}{3} \gamma_{ij} \Delta
  \right) 
  \left( \Delta + 3 K \right)^{-1}
  \left(
    \Delta^{-1} D^{k}D^{l}\Gamma_{kl}
    - \frac{1}{3} \Gamma_{k}^{\;\;k}
  \right)
  \nonumber\\
  &&
  - 4
  \left( 
    D_{(i}\left(\Delta+2K\right)^{-1}D_{j)}\Delta^{-1}D^{l}D^{k}\Gamma_{lk}
    - D_{(i}\left(\Delta+2K\right)^{-1}D^{k}\Gamma_{j)k}
  \right)
  = 0
  ,
  \label{eq:kouchan-19.124}
\end{eqnarray}
where we denote $\Gamma_{i}^{\;\;j} = \gamma^{jk}\Gamma_{ik}$.
In these equations, $\stackrel{(2)}{{\cal E}}$ and
$\stackrel{(2)}{{\cal P}}$ are the second-order perturbations of the
energy density and the pressure, respectively.
Further, $D_{i}\stackrel{(2)}{v}$ and 
$\stackrel{(2)}{{\cal V}_{i}}$ are the scalar- and the
vector-parts of the spatial components of the covariant fluid
four-velocity, in these equations.
$\Gamma_{0}$, $\Gamma_{i}$, and $\Gamma_{ij}$ are the
collections of the quadratic terms of the linear-order
perturbations in the second-order Einstein equations and these
can be regarded as the source terms in the second-order Einstein
equations.
The explicit form of these source terms are given in
Refs.~\cite{kouchan-JGRG17-proceedigns,kouchan-second-cosmo-consistency}. 
First-order perturbations of the Einstein equations are
given by the replacements
$\stackrel{(2)}{\Phi}\rightarrow\stackrel{(1)}{\Phi}$, 
$\stackrel{(2)}{\Psi}\rightarrow\stackrel{(1)}{\Psi}$, 
$\stackrel{(2)}{\nu_{i}}\rightarrow\stackrel{(1)}{\nu_{i}}$, 
$\stackrel{(2)}{\chi_{ij}}\rightarrow\stackrel{(1)}{\chi_{ij}}$, 
$\stackrel{(2)}{{\cal E}}\rightarrow\stackrel{(1)}{{\cal E}}$, 
$\stackrel{(2)}{{\cal P}}\rightarrow\stackrel{(1)}{{\cal P}}$, 
$D_{i}\stackrel{(2)}{v}\rightarrow D_{i}\stackrel{(1)}{v}$, 
$\stackrel{(2)}{{\cal V}_{i}}\rightarrow\stackrel{(1)}{{\cal V}_{i}}$,
and $\Gamma_{0}=\Gamma_{i}=\Gamma_{ij}=0$.


\section{Consistency with the equations of motion for matter field}
\label{sec:Consystency-with-the-equation-second}


Now, we consider the second-order perturbation of the energy
continuity equation and the Euler equations.
In terms of gauge-invariant variables, the second-order
perturbations of the energy continuity equation and the Euler
equation for a single perfect fluid are given
by\cite{kouchan-second-cosmo-matter}
\begin{eqnarray}
  a {}^{(2)}\!{\cal C}_{0}^{(p)}
  \!\!\!&:=&\!\!\!
      \partial_{\eta}\stackrel{(2)}{{\cal E}} 
  + 3 {\cal H}
  \left(
      \stackrel{(2)}{{\cal E}} 
    + \stackrel{(2)}{{\cal P}}
  \right)
  + \left(\epsilon + p\right) \left(
         \Delta \stackrel{(2)}{v}
    -  3 \partial_{\eta}\stackrel{(2)}{\Psi}
  \right)
  - \Xi_{0}
  = 0
  \label{eq:kouchan-19.132}
  , \\
  {}^{(2)}\!{\cal C}_{i}^{(p)}
  \!\!\!&:=&\!\!\!
  \left( \epsilon + p \right) \left\{
    \left(
      \partial_{\eta} + {\cal H}
    \right)
    \left(
      D_{i}\stackrel{(2)}{v}
      + \stackrel{(2)}{{\cal V}_{i}}
    \right)
    + D_{i}\stackrel{(2)}{\Phi}
  \right\}
  + D_{i}\stackrel{(2)}{{\cal P}}
  + \partial_{\eta}p \left(
      D_{i}\stackrel{(2)}{v}
    + \stackrel{(2)}{{\cal V}_{i}}
  \right)
  - \Xi_{i}^{(p)}
  = 0,
  \label{eq:kouchan-17.680}
\end{eqnarray}
where $\Xi_{0}$ and $\Xi_{i}^{(p)}$ are the collection of the
quadratic terms of the linear order perturbations and its
explicit forms are given in
Ref.~\cite{kouchan-second-cosmo-matter,kouchan-second-cosmo-consistency}.


To confirm the consistency of the background and the
perturbations of the Einstein equation and the energy continuity
equation (\ref{eq:kouchan-19.132}), we first substitute the
second-order Einstein equations
(\ref{eq:background-Einstein-equation-1})--(\ref{eq:kouchan-19.121}),
and (\ref{eq:kouchan-19.125}) into Eq.~(\ref{eq:kouchan-19.132}).
For simplicity, we first impose the first-order version of
Eq.~(\ref{eq:kouchan-19.122}) on all equations.
Then, we obtain
\begin{eqnarray}
  4 \pi G a^{3} {}^{(2)}\!{\cal C}_{0}^{(p)}
  &=&
  -             \partial_{\eta}\stackrel{(2)}{{}^{(p)}E_{(1)}}
  -             {\cal H} \stackrel{(2)}{{}^{(p)}E_{(1)}}
  -          3  {\cal H} \stackrel{(2)}{{}^{(p)}E_{(2)}}
  +             D^{i}\stackrel{(2)}{{}^{(p)}E_{(4)i}}
  + \frac{3}{2} \left(
    3 \stackrel{(0)}{{}^{(p)}E_{(1)}}
    - \stackrel{(0)}{{}^{(p)}E_{(2)}}
  \right) \partial_{\eta}\stackrel{(2)}{\Psi}
  \nonumber\\
  &&
  -             \partial_{\eta}\Gamma_{0}
  -             {\cal H} \Gamma_{0}
  - \frac{1}{2} {\cal H} \Gamma_{k}^{\;\;k}
  + \frac{1}{2} D^{k}\Gamma_{k}
  - 4 \pi G a^{2} \Xi_{0}
  .
  \label{eq:kouchan-continuity-consistency}
\end{eqnarray}
This equation shows that the second-order perturbation
(\ref{eq:kouchan-19.132}) of the energy continuity equation is
consistent with the second-order and the background Einstein
equations if the equation 
\begin{eqnarray}
    4 \pi G a^{2} \Xi_{0}
  +   \left(
    \partial_{\eta}
    +   {\cal H}
  \right) \Gamma_{0}
  + \frac{1}{2} {\cal H} \Gamma_{k}^{\;\;k}
  - \frac{1}{2} D^{k}\Gamma_{k}
  =
  0
  \label{eq:kouchan-19.135}
\end{eqnarray}
is satisfied under the background, the first-order Einstein
equations.
Actually, through the background Einstein equations
(\ref{eq:background-Einstein-equation-1}) and the first-order
version of the Einstein equations 
(\ref{eq:kouchan-19.120})--(\ref{eq:kouchan-19.124}), we can 
easily see that Eq.~(\ref{eq:kouchan-19.135}) is satisfied under
the Einstein equations of the background and of the first
order\cite{kouchan-second-cosmo-consistency}.


Next, we consider the second-order perturbations of the Euler
equations.
For simplicity, we first impose the first-order version of
Eq.~(\ref{eq:kouchan-19.122}) on all equations, again.
Through the background Einstein equations
(\ref{eq:background-Einstein-equation-1}) and the Einstein
equations of the second order 
(\ref{eq:kouchan-19.121})--(\ref{eq:kouchan-19.125}), we can 
obtain
\begin{eqnarray}
  8 \pi G a^{2} {}^{(2)}\!{\cal C}_{i}^{(p)}
  &=&
  -  8 \pi G a^{3} \stackrel{(0)}{C_{0}^{(p)}} \left(
    D_{i}\stackrel{(2)}{v} + \stackrel{(2)}{{\cal V}_{i}} 
  \right)
  -    D_{i}\stackrel{(2)}{\Phi} \left(
    3 \stackrel{(0)}{{}^{(p)}E_{(1)}}
    - \stackrel{(0)}{{}^{(p)}E_{(2)}}
  \right)
  -  2 D_{i}\stackrel{(2)}{{}^{(p)}E_{(2)}}
  \nonumber\\
  &&
  - \frac{2}{3} D_{i}\left(
    \Delta + 3 K
  \right) \stackrel{(2)}{{}^{(p)}E_{(3)}}
  + \frac{1}{2}
  \left(
    \partial_{\eta} + 2 {\cal H}
  \right) 
  \left(
    + 4 \stackrel{(2)}{{}^{(p)}E_{(4)i}}
    -   \stackrel{(2)}{{}^{(p)}E_{(5)i}}
  \right)
  \nonumber\\
  &&
  + \frac{1}{2a^{2}} \left(\Delta + 2 K\right)\stackrel{(2)}{{}^{(p)}E_{(6)i}}
  - 8 \pi G a^{2} \Xi_{j}^{(p)} 
  + \left(
    \partial_{\eta} + 2 {\cal H}
  \right) \Gamma_{j}
  - D^{l}\Gamma_{jl}
  .
  \label{eq:kouchan-19.150}
\end{eqnarray}
This equation shows that the second-order perturbations of the
Euler equations is consistent with the Einstein equations of the
background and the second order if the equation
\begin{eqnarray}
  \left(
    \partial_{\eta} + 2 {\cal H}
  \right) \Gamma_{j}
  -    D^{l}\Gamma_{jl}
  -  8 \pi G a^{2} \Xi_{j}^{(p)} 
  &=& 0
  \label{eq:kouchan-19.152}
\end{eqnarray}
is satisfied under the Einstein equations of the background and
the first order.
Actually, we can easily confirm Eq.~(\ref{eq:kouchan-19.152})
due to the background Einstein equations and the first-order
perturbations of the Einstein
equations\cite{kouchan-second-cosmo-consistency}, and implies
that the second-order perturbation of the Euler equation is
consistent with the set of the background, the first-order, and
the second-order Einstein equations.


The consistency of equations for perturbations shown here is
just a well-known result, i.e., the Einstein equation includes
the equations of motion for matter field due to the Bianchi
identity.
However, the above verification of the identities
(\ref{eq:kouchan-19.135}) and (\ref{eq:kouchan-19.152}) implies
that our derived second-order perturbations of the Einstein
equation, the equation of continuity, and the Euler equation are
consistent.
In this sense, we may say that the derived second-order Einstein
equations, especially, the derived formulae for the source terms
$\Gamma_{0}$, $\Gamma_{i}$, $\Gamma_{ij}$, $\Xi_{0}$, and
$\Xi_{i}$ in Ref.~\cite{kouchan-second-cosmo-consistency} are
correct.


\section{Summary}
\label{sec:summary}


In summary, we show the all components of the second-order
perturbation of the Einstein equation without ignoring any modes
of perturbation in the case of a perfect fluid.
The derivation is based on the general framework of the
second-order gauge-invariant perturbation theory developed in
Refs.~\cite{kouchan-gauge-inv}.
In this formulation, any gauge fixing is not necessary and we
can obtain any equation in the gauge-invariant form which is
equivalent to the complete gauge fixing.
In other words, our formulation gives complete gauge fixed
equations without any gauge fixing.
Therefore, equations which are obtained in gauge-invariant
manner cannot be reduced without physical restrictions any more. 
In this sense, these equations are irreducible.
This is one of the advantages of the gauge-invariant perturbation
theory.


We have also checked the consistency of the set of equations of
the second-order perturbation of the Einstein equations and the
evolution equation of the matter field in the cases of a perfect
fluid.
Therefore, in the case of the single matter field, we may say
that we have been ready to clarify the physical behaviors of the
second-order cosmological perturbations.
The physical behavior of the second-order perturbations in the
universe filled with a single matter field will be instructive
to clarify those of the second-order perturbations in more
realistic cosmological situations.
We leave these issues as future works.



\end{document}